\documentclass[conference,compsoc]{IEEEtran}
\usepackage[cmex10]{amsmath}
\usepackage{graphicx,marvosym}
\usepackage{setspace}
\usepackage{listings}
\usepackage{multicol} 
\usepackage{subfigure}
\usepackage{amssymb}
\ifCLASSINFOpdf
\else
\fi
\begin{document}

\interdisplaylinepenalty=2500

\lstset{language=C,basicstyle=\small}
\lstset{numbers=left, numberstyle=\tiny, stepnumber=1, numbersep=5pt}
\lstset{tabsize=2}
\lstset{firstnumber=1}
\lstset{frame=single}
\lstset{
  language={C},
  morekeywords={assert,uchar}
}

%
\title{SMT-Based Bounded Model Checking for Embedded ANSI-C Software}


\author{\IEEEauthorblockN{Lucas Cordeiro}
\IEEEauthorblockA{University of Southampton\\
lcc08r@ecs.soton.ac.uk}
\and
\IEEEauthorblockN{Bernd Fischer}
\IEEEauthorblockA{University of Southampton\\
b.fischer@ecs.soton.ac.uk}
\and
\IEEEauthorblockN{Joao Marques-Silva}
\IEEEauthorblockA{University College Dublin\\
jpms@ucd.ie}}


%


\maketitle

\begin{abstract}
Propositional bounded model checking has been applied successfully to verify embedded software but is limited by the increasing propositional formula size and the loss of structure during the translation. These limitations can be reduced by encoding word-level information in theories richer than propositional logic and using SMT solvers for the generated verification conditions. Here, we investigate the application of different SMT solvers to the verification of embedded software written in ANSI-C. We have extended the encodings from previous SMT-based bounded model checkers to provide more accurate support for finite variables, bit-vector operations, arrays, structures, unions and pointers. We have integrated the CVC3, Boolector, and Z3 solvers with the CBMC front-end and evaluated them using both standard software model checking benchmarks and typical embedded applications from telecommunications, control systems and medical devices. The experiments show that our approach can analyze larger problems and substantially reduce the verification time.
\end{abstract}


%
\IEEEpeerreviewmaketitle

\section{Introduction}
\label{01-introduction}

Bounded Model Checking (BMC) based on Boolean Satisfiability (SAT) has been introduced as a complementary technique to Binary Decision Diagrams (BDD's) for alleviating the state explosion problem~\cite{Biere99}. The basic idea of BMC is to check (the negation of) a given property at a given depth: given a transition system \textit{M}, a property $\phi$, and a bound \textit{k}, BMC unrolls the system \textit{k} times and translates it into a verification condition $\psi$ such that $\psi$ is satisfiable if and only if $\phi$ has a counterexample of depth less than or equal to \textit{k}. Standard SAT checkers can be used to check if $\psi$ is satisfiable. In order to cope with increasing system complexity, SMT (Satisfiability Modulo Theories) solvers can be used as back-ends for solving the verification conditions generated from the BMC instances~\cite{Armando06,Ganai06,Xu08,Armando09}. In SMT, predicates from various (decidable) theories are not encoded by propositional variables as in SAT, but remain in the problem formulation. These theories are handled by dedicated decision procedures. Thus, in SMT-based BMC, $\psi$ is a quantifier-free formula in a decidable subset of first-order logic which is then checked for satisfiability by an SMT solver.



In order to reason about embedded software accurately, an SMT-based BMC must consider a number of issues that are not easily mapped into the theories supported by SMT solvers. In previous work on SMT-based BMC for software~\cite{Armando06,Ganai06,Armando09} only the theories of uninterpreted functions, arrays and linear arithmetic were considered, but no encoding was provided for ANSI-C~\cite{ISO99} constructs such as bit operations, floating-point arithmetic, pointers (e.g., pointer arithmetic and comparisons) and unions. This limits its usefulness for analyzing and verifying embedded software written in ANSI-C. In addition to that, the SMT-based BMC approach proposed by Armando et al.~\cite{Armando06,Armando09} does not support the checking of arithmetic overflow and does not make use of high-level information to simplify the unrolled formula. We address these limitations by exploiting the different background theories of SMT solvers to build an SMT-based BMC tool that precisely translates program expressions into quantifier-free formulae and applies a set of optimization techniques to prevent overburdening the solver. This way we achieve significant performance improvements over SAT-based BMC and previous work on SMT-Based BMC.




This work makes two major novel contributions. First, we provide details of an accurate translation from ANSI-C programs into quantifier-free formulae. Second, we demonstrate that the new approach improves the performance of software model checking for a wide range of embedded software systems. Additionally, we show that our encoding allows us to reason about arithmetic overflow and to verify programs that make use of bit-level, pointers, unions and floating-point arithmetic. We also use three different SMT solvers (CVC3, Boolector and Z3) in order to check the effectiveness of our encoding techniques. To the best of our knowledge, this is the first work that reasons accurately about ANSI-C constructs commonly found in embedded software and extensively applies SMT solvers to check the verification conditions emerging from the BMC of embedded software industrial applications. We describe the ESW-CBMC tool that extends the C Bounded Model Checker (CBMC)~\cite{Clarke04} to support different SMT solvers in the back-end and to make use of high-level information to simplify and reduce the unrolled formula size. Experimental results obtained with ESW-CBMC show that our approach scales significantly better than both the CBMC model checker~\cite{Clarke04} and SMT-CBMC, a bounded model checker for C programs that is based on the SMT solvers CVC3 and Yices~\cite{Armando06,Armando09}.



\section{Background}
\label{02-background}

ESW-CBMC uses the front-end of CBMC to generate the verification conditions (VCs) for a given ANSI-C program. However, instead of passing the VCs to a propositional SAT solver, we convert them using different background theories and pass them to an SMT solver. In this section, we describe the main features of CBMC and present the background theories used in the rest of the paper.

\subsection{C Bounded Model Checker}
\label{02-CBoundedModelChecker}

CBMC implements BMC for ANSI-C/C++ programs using SAT solvers~\cite{Clarke04}. It can process C/C++ code using the goto-cc tool~\cite{Wintersteiger09}, which compiles the C/C++ code into equivalent GOTO-programs using a gcc-compliant style. Alternatively, CBMC uses its own, internal parser based on Flex/Bison, to process the C/C++ files and to build an abstract syntax tree (AST). The typechecker annotates this AST with types and generates a symbol table. CBMC's IRep class then converts the annotated AST and the C/C++ GOTO-programs into an internal, language-independent format used by the remaining phase of the front-end. 

CBMC derives the VCs using two recursive functions that compute the \textit{assumptions} or \textit{constraints} (i.e., variable assignments) and \textit{properties} (i.e., safety conditions and user-defined assertions). CBMC's VC generator (VCG) automatically generates safety conditions that check for arithmetic overflow and underflow, array bounds violations and null-pointer dereferences. Both functions accumulate the control flow predicates to each program point and use that to guard both the constraints and the properties, so that they properly reflect the program's semantics. 

Although CBMC implements several state-of-the-art techniques for propositional BMC, it still has the following limitations~\cite{Armando06,Ganai06}: \textit{(i)} large data-paths involving complex expressions lead to large propositional formulae, \textit{(ii)} high-level information is lost when the verification conditions are converted into propositional logic, and \textit{(iv)} size of the encoding increases with the size of the arrays used in the program.

\subsection{Satisfiability Modulo Theories}
\label{02-SatisfiabilityModuloTheories}

SMT decides the satisfiability of first-order formulae using the combination of different background theories and thus generalizes propositional satisfiability by supporting uninterpreted functions, arithmetic, bit-vectors, tuples, arrays, and other decidable first-order theories. SMT solvers are decision procedures for certain theories: given a decidable theory ${\cal T}$ and a quantifier-free formula $\psi$, they check whether $\psi$ is satisfiable in ${\cal T}$ or not, or equivalently, whether ${\cal T}\cup \{\psi\}$ is satisfiable. Given a set $\Gamma\cup \{\psi\}$ of formulae over $\cal T$, we say that $\psi$ is a $\cal T$-consequence of $\Gamma$, and write $\Gamma\models_{\cal T}\psi$, if and only if every model of ${\cal T}\cup\Gamma$ is also a model of $\psi$. Checking $\Gamma\models_{\cal T}\psi$ can be reduced in the usual way to checking the $\cal T$-satisfiability of $\Gamma\cup\{\neg\psi\}$. 

In SMT-based bounded model checking, we unroll the transition system $M$ and the property $\psi$ (which is to be checked in ${\cal T}$), yielding $M$$_{k}$ and $\psi$$_{k}$ respectively, and pass these to an SMT solver to check $M_{k}\models_{\cal T}\psi_{k}$~\cite{Ganai06}. The solver will always terminate with a satisfiable/unsatisfiable answer. If the answer is satisfiable, we have found a violation of the property $\psi$. If it is unsatisfiable, the property $\psi$ holds in $M$ up to the given bound $k$.

State-of-the-art SMT solvers support not only the combination of different decidable theories, but also the integration of SAT solvers in order to speed up the performance. Furthermore, they often also integrate a simplifier which applies standard algebraic reduction rules before \textit{bit-blasting} (i.e., replacing the word-level operators by bit-level circuit equivalents) propositional expressions to a SAT solver. Background theories vary but the SMT-LIB initiative aims at establishing a common standard for the specification of background theories~\cite{smtlib09}. However, most SMT solvers provide functions in addition to those specified in the SMT-LIB. Therefore, we describe here all the fragments that we found in the SMT solvers CVC3, Boolector and Z3 for the theory of linear, non-linear, and bit-vector arithmetic~\cite{CVC07,Boolector09,Z08}. We summarize the syntax of these background theories as follows:

\[\begin{array}{rcl}
\mathit{Fml}  & ::= & \mathit{Fml} \: \mathit{con} \: \mathit{Fml} \:
                    | \: \neg\mathit{Fml} \:
                    | \: Atom \\
\mathit{con}  & ::= & \: \wedge \: 
                    | \: \vee \: 
                    | \: \oplus \:
                    | \: \Rightarrow \:
                    | \: \Leftrightarrow \: \\                    
\mathit{Atom} & ::= &  \mathit{Trm} \: \mathit{rel} \: \mathit{Trm} \:
                    | \: \mathit{Id} \: | \: true \: | \: false \\
\mathit{rel}  & ::= & \: < \: 
                    | \: \leq \: 
                    | \: > \: 
                    | \: \geq
                    | \: = \: 
                    | \: \neq \: \\
\mathit{Trm}  & ::= &  \mathit{Trm} \: op \: \mathit{Trm} \: 
                    | \: \mathit{Const} \: 
                    | \: \mathit{Id} \: 
                    | \: \mathit{Extract}\left[i,j\right] \\
              &   & | \: \mathit{SignExt}\left[k\right]
                    | \: \mathit{ZeroExt}\left[k\right] \\
              &   & | \: \mathit{ite}\left(\mathit{Fml}, \: \mathit{Trm}, \mathit{Trm}\right) \: \\                 
\mathit{op}   & ::= & +_{o,u} \: 
                    | \: -_{o,u} \: 
                    | \: *_{o,u} \: 
                    | \: /_{o} \: 
                    | \: rem \: \\ 
              &   & | \: << \: 
                    | \: >> \: 
                    | \: \& \: 
                    | \: 
                    | \: 
                    | \: \oplus \: 
                    | \: @ \:  
\end{array}\]

\noindent In this grammar \emph{Fml} denotes Boolean-valued expressions, \emph{Trm} denotes integers, reals, and bit-vectors while \emph{op} denotes binary operators. The semantics of the relational operators (i.e., $<$, $\leq$, $>$, $\geq$), the non-linear arithmetic operators (i.e., $*$, $/$, \emph{rem}) and the right-shift operator ($>>$) depends on whether the program variables are unsigned or signed bit-vectors, integers or real numbers. The expression $\mathit{Extract}\left[i,j\right]$ denotes bit-vector extraction from bits \textit{i} down to \textit{j} to yield a new bit-vector of size  $i-j+1$ while $@$ denotes the concatenation of the given bit-vectors. $\mathit{SignExt}\left[k\right]$ extends the bit-vector to the signed equivalent bit-vector of size $w+k$, where $w$ is the original width of the bit-vector, while $\mathit{ZeroExt}\left[k\right]$ extends the bit-vector with zeros to the unsigned equivalent bit-vector of size $w+k$. The conditional expression $\mathit{ite}\left(\mathit{Fml}, \: \mathit{Trm}, \mathit{Trm}\right)$ takes as first argument a Boolean formula and depending on its value, selects either the second or the third argument. The indexes $o$ and $u$ in the operators $+$, $-$, $*$ and $/$ denote predicates that check if the bit-wise addition, subtraction, multiplication and division overflow and underflow respectively. The operator \emph{rem} denotes the signed or unsigned remainder. 

The array theories of SMT solvers are typically based on the two McCarthy axioms~\cite{McCarthy62}. Let \textit{a} be an array, \textit{i} and \textit{j} be integers and \textit{v} be a value. The function \textit{select(a,i)} denotes the value of array \textit{a} at index \textit{i} and \textit{store(a,i,v)} denotes an array that is exactly the same as array \textit{a} except that the value at index position \textit{i} is \textit{v} (if \textit{i} is within the array bounds). Formally, the functions \textit{select(a,i)} and \textit{store(a,i,v)} can then be represented by the following two axioms~\cite{CVC07,Boolector09,Z08}:
\begin{eqnarray}
  \left(i=j \Rightarrow select\left(store\left(a,i,v\right),j\right)=v\right) \nonumber \\
  \left(i \neq j \Rightarrow select\left(store\left(a,i,v\right),j\right)=select\left(a,j\right)\right) \nonumber
\end{eqnarray}

\noindent
The first axiom asserts that the value selected at index \textit{j} is the same as the last value stored to the index \textit{i}, if the two indices \textit{i} and \textit{j} are equal. The second axiom asserts that storing a value to index \textit{i}, does not change the value at index \textit{j}, if the indices \textit{i} and \textit{j} are different. 

Tuples are used to model the ANSI-C unions and struct datatypes. They provide \textit{store} and \textit{select} operations similar to those in arrays, but working on the tuple \textit{elements}. Hence, the expression \textit{select(t,f)} denotes the field \textit{f} of tuple \textit{t} while the expression \textit{store(t,i,v)} denotes that tuple \textit{t} at field \textit{f} has the value \textit{v} and all other tuple elements remain the same.

\section{ESW-CBMC}
\label{03-SMT-Based-CBMC}

This section describes the main software components that are integrated into the SMT-based back-end of CBMC and the encoding techniques that we used to convert the constraints and properties from the ANSI-C embedded software into the background theories of the SMT solvers.

\subsection{SMT-based CBMC Back-End}
\label{sec:TheNewBackEndOfCBMC}

\begin{figure}[ht]
\centering
\includegraphics[scale=0.20]{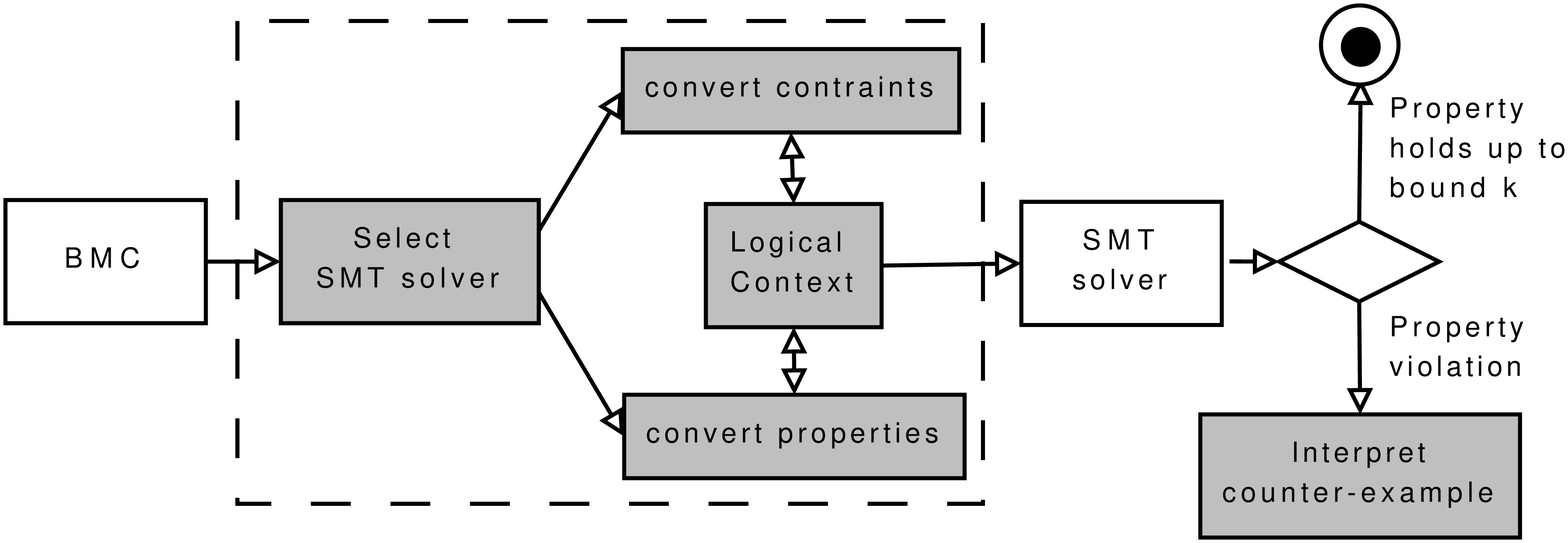}
\caption{Overview of the SMT-based CBMC Back-end.}
\label{figure:system-architecture}
\end{figure}

Figure~\ref{figure:system-architecture} shows the new back-end of CBMC in order to support the SMT solvers CVC3, Boolector and Z3. The gray boxes represent the components that we modified/included in the back-end of CBMC. We reused the front-end completely unchanged, i.e., we process the constraints and properties that CBMC's VCG generates for the unrolled C program in single static assignment (SSA) form. However, we implemented a new pair of encoding functions for each supported SMT solver and let the user select between them. The selected functions are then used to encode the given constraints and properties into a global logical context, using the background theories supported by the selected SMT solver. Finally, we invoke this solver to check the satisfiability of the context formula.

Formally, we build two sets of quantifier-free formulae $C$ (for the constraints) and $P$ (for the properties) such that $M$$\not\models$$_{k}$$C$$\wedge$$\neg$$P$ if and only if the property $P$ holds in the model $M$ up to the bound $k$. If not, we have found a violation of the property $P$. However, this approach can be used only to find violations of the property up to the bound $k$ and not to prove properties. For software verification, in order to \emph{prove} properties we need to compute the \textit{completeness threshold} to determine the maximum number of loop-iterations occurring in the program~\cite{Ganai08,handbook09}. Worst-case execution time (WCET) tools can be used to compute the \textit{completeness threshold} by means of static analysis of loop structures. The WCET essentially indicates the maximum number of loop-iterations and as a result CBMC and ESW-CBMC adopt this approach to compute the maximum bound of the program. However, in practice, complex software programs involve large data-paths and complex expressions. Therefore, the resulting formulae become harder to solve and require substantial amounts of memory to build. Thus, for complex software programs, we can only ensure that the property holds in $M$ up to a bound $k$.


We use the code in Figure~\ref{fig:set-of-transformation} as a running example to illustrate the process to transform a given ANSI-C code into SSA form and after that into the quantifier-free formulae $C$ and $P$ (as shown in (\ref{quantifier-free-formulae-c}) and (\ref{quantifier-free-formulae-p})). It is important to note that the code of Figure~\ref{fig:set-of-transformation}(a) is a syntactically valid C program, but it writes accidentally to an address outside the allocated memory region of the array \emph{a} (line 6). Hence, in order to reason about this C program, seven VCs are generated as follows: the first six VCs check the lower and upper bound of array \emph{a} in lines 4, 6 and 7 respectively and the last VC checks the \emph{assert} macro defined by the user in line 7. However, before actually checking the properties, the front-end of CBMC performs a set of transformations and converts the program into SSA form. As a result, the original C program in Figure~\ref{fig:set-of-transformation}(a) is then converted into SSA form that only consists of \emph{if} instructions, assignments and assertions as shown in Figure~\ref{fig:set-of-transformation}(b).

\begin{figure}[ht]
\begin{lstlisting}
int main() { 
int a[2], i, x;
if (x==0)
  a[i]=0;
else
  a[i+2]=1;
assert(a[i+1]==1);
}
\end{lstlisting}
\centerline{(a)}
\begin{lstlisting}
g1 == (x1 == 0) 
a1 == (a0 WITH [i0:=0])
a2 == a0
a3 == (a2 WITH [2+i0:=1])
a4 == (g1 ? a1 : a3)
t1 == (a4[1+i0] == 1)
\end{lstlisting}
\centerline{(b)}
\caption{(a) A C program with violated property. (b) The C program of (a) in SSA form.}
\label{fig:set-of-transformation} 
\end{figure}
\begin{equation}
\label{quantifier-free-formulae-c}
C := \left [ \begin{array}{ll} 
                g_{1} := (x_{1} = 0) \\ 
                \wedge \, \, a_{1} := store(a_{0}, i_{0}, 0) \\
                \wedge \, \, a_{2} := a_{0} \\
                \wedge \, \, a_{3} := store(a_{2}, 2+i_{0}, 1) \\
                \wedge \, \, a_{4} := ite(g_{1}, a_{1}, a_{3})
              \end{array} \right ]  \\ 
\end{equation}
\begin{equation}              
\label{quantifier-free-formulae-p}
P := \left [ \begin{array}{ll} 
                i_{0} \geq 0 \wedge i_{0} < 2 \\ 
                \wedge \, \, 2+i_{0} \geq 0 \wedge 2+i_{0} < 2 \\
                \wedge \, \, 1+i_{0} \geq 0 \wedge 1+i_{0} < 2\\
                \wedge \, \, select(a_{4}, i_{0}+1) = 1 \\
              \end{array} \right ]              
\end{equation}

From this, we build the constraints and properties formulae shown in (\ref{quantifier-free-formulae-c}) and (\ref{quantifier-free-formulae-p}). We use additional boolean variables (called definition \textit{literals}) for each clause of the formula $P$ in such a way that the definition literal is true \textit{iff} a given clause of the formula $P$ is true. Hence, in the example we add a constraint for each clause of $P$ as shown in (\ref{quantifier-free-formulae-p-p}):
\begin{eqnarray}
\label{quantifier-free-formulae-p-p}
l_{0} \Leftrightarrow i_{0} \geq 0 \nonumber \\ 
l_{1} \Leftrightarrow i_{0} < 2 \nonumber \\ 
\vdots \nonumber \\ 
l_{6} \Leftrightarrow select(a_{4}, i_{0}+1) = 1
\end{eqnarray}

\noindent We then rewrite (\ref{quantifier-free-formulae-p}) as:
\begin{eqnarray}
\label{quantifier-free-formulae-p-p-p}
\neg P := \neg l_{0} \vee \neg l_{1} \vee \ldots \vee \neg l_{6}
\end{eqnarray}

\noindent It is also important to point out that we simplify the formulae $C$ and $P$ by using local and recursive transformations in order to remove functionally redundant expressions and redundant literals. Finally, the resulting formula $C$$\wedge$$\neg$$P$ is passed to an SMT solver to check satisfiability. This is different to the approach by Armando et al.~\cite{Armando06,Armando09} who build two sets of quantifier-free formulae ${\cal C}$ and ${\cal P}$ and check whether ${\cal C}\models_{\cal T}\bigwedge{\cal P}$ using an SMT solver. Moreover, they transform the C code into conditional normal form instead of single static assignment form as we do in this work.

As mentioned in Section~\ref{02-SatisfiabilityModuloTheories}, modern SMT solvers provide ways to model the program variables as bit-vectors or as elements of a numerical domain (e.g., $\mathbb{Z}$, $\mathbb{R}$). If the program variables are modelled as bit-vectors of fixed size, then the result of the analysis can be precise (w.r.t. the ANSI-C semantics) depending on the size considered for the bit-vectors. On the other hand, if the program variables are modelled as numerical values, then the result of the analysis is independent from the actual binary representation, but the analysis may not be precise when arithmetic expressions are involved. For instance, the following formula is valid in numerical domains such as $\mathbb{Z}$ or $\mathbb{R}$:
\begin{equation}
\left(a > 0 \wedge b > 0\right) \Rightarrow \left(a + b > 0\right)
\end{equation}

\noindent However, it does not hold if \textit{a} and \textit{b} are interpreted as bit-vectors of fixed-size, due to possible overflow in the addition operation (Section~\ref{sec:Encoding} explains how we encode arithmetic overflow). In our benchmarks, we noted that the majority of VCs are solved faster if we model the basic datatypes as integer and/or real. Therefore, we have to trade off \textit{speed} and \textit{accuracy} which might be two competing goals in formal verification using SMT solvers. Speed results from the omission of detail in the original C program, whereas accuracy results from the inclusion of detail. When encoding the constraints and properties of C programs into SMT, we allow the verification engineer to decide the way to model the basic data types (i.e., as integer/real values or as bit-vectors) through a run-time option of ESW-CBMC.

\subsection{Code Optimizations}
\label{sec:CodeOptimizations}

The ESW-CBMC tool implements some standard code optimization techniques such as constant folding and forward substitution~\cite{Muchnick97}. We observe that there is a representative number of embedded applications in which these optimization techniques make a significant impact on the performance of the tool. Constant folding, which is implemented in the front-end allows us to replace arithmetic operations involving constants by other constants that represent the result of the operation. Figure~\ref{figure:code-fragment-of-cyclic-redundancy-check} shows an example of constant folding when applied to the cyclic redundancy check algorithm extracted from the SNU Real-Time benchmark~\cite{SNURealTime09}.

\begin{figure}[ht]
\centering
\begin{lstlisting}
for (j=0;j<=255;j++) {
  icrctb[j]=icrc1(j<<8,(uchar)0);
	rc[j]=(uchar)(it[j&0xF]<<4 | it[j>>4]);
}
\end{lstlisting}
\caption{Code fragment of cyclic redundancy check.}
\label{figure:code-fragment-of-cyclic-redundancy-check}
\end{figure}

The right hand side of the expressions in line 2 and 3 are replaced by the corresponding constants since the value of the variable \textit{j} and all elements of array \textit{it} (where \textit{it} is an array of constants) are known at verification time. As a result, we can encode the expressions in line 2 and 3 by using only the function \textit{store} of the SMT solvers (note that the function \textit{icrc1} receives two arguments and returns another element of type unsigned char). We also observed that there are several embedded applications that repeat the same expression many times at different places. The value of the operands in the expression does not change in between the two evaluations of that expression and can thus be forward substituted. Figure~\ref{figure:code-fragment-of-fast-fourier-transform} shows an example of the forward substitution technique when applied to the Fast Fourier Transform algorithm extracted again from the SNU Real-Time benchmark~\cite{SNURealTime09}.

\begin{figure}[ht]
\centering
\begin{lstlisting}
typedef struct {
  float real, imag;
} complex;
complex x[1024], *xi;
for(le=n/2; le>0; le/=2) {
  ...
  for (j=0; j<le; j++) {
    ...
    for (i=j; i<n; i=i+2*le) {
      xi = x + i;
      ...
    }
  }
}
\end{lstlisting}
\caption{Code fragment of fast fourier transform.}
\label{figure:code-fragment-of-fast-fourier-transform}
\end{figure}

The right hand side of the assignment in line 10 is repeated according to the bound used to model check this program. This occurs because the most outer \emph{for} loops (lines 5-14 and lines 7-13) invoke the most inner \emph{for} loop (lines 9-12) \textit{n} times (where \textit{n} represents the unwinding bound) and the address of the array \textit{x} also does not change inside the loops. For instance, if the bound is set to 1024, then the expression \textit{x + i} that is assigned to the \textit{xi} pointer index is repeated 1024 times (note that this expression involves pointer arithmetic). As a result, we include all expressions into a cache so that when a given expression is processed again in the program, we only retrieve it from the cache instead of creating a new set of variables.

\subsection{Encodings}
\label{sec:Encoding}

\subsubsection{Scalar Data Types}
\label{03-EncodingScalarDataTypes}

We provide two approaches to model unsigned and signed integer data types, either as the integers provided by the corresponding SMT-lib theories or as bit-vectors, which are encoded using a particular bit width such as 32 bits. The relational operators (e.g., $<$, $\leq$, $>$, $\geq$), arithmetic operators (e.g., $+$, $-$, $/$, $*$, \emph{rem}) and right-shift are encoded depending on whether the operands are unsigned or signed bit-vectors, integer or real numbers. We support all type casts, including conversion between integer and floating-point types. From the front-end's point of view, there are six scalar datatypes: \emph{bool}, \emph{signedbv}, \emph{unsignedbv}, \emph{fixedbv}, \emph{floatbv} and \emph{pointer}. At this point in time, we only support fixed-point arithmetic (i.e., \emph{fixedbv}) for \emph{double and float} instead of floating-point arithmetic (i.e., \emph{floatbv}). 

The ANSI-C datatypes \emph{int, long int, long long int, char} are considered as \emph{signedbv} with different bit width (depending on the machine architecture) and the unsigned version of these datatypes are considered as \emph{unsignedbv}. The conversions between \emph{signedbv}, \emph{unsignedbv} and \emph{fixedbv} are performed using the world-level functions $\mathit{Extract}\left[i,j\right]$, $\mathit{SignExt}\left[k\right]$ and $\mathit{ZeroExt}\left[k\right]$ (described in Section~\ref{02-SatisfiabilityModuloTheories}). Similarly, upon dereferencing, the object that the pointer points to is converted using the same word-level functions. The datatype \emph{bool} is converted into \emph{signedbv} and \emph{unsignedbv} using $\mathit{ite}$. In addition, \emph{signedbv} and \emph{unsignedbv} are converted into \emph{bool} using the operator $\neq$ by comparing the variable to be converted with zero. Formally, let \textit{v} be a variable of signed type, \textit{k} be a constant whose value is zero matching the type of \textit{v} and let \textit{t} be a boolean variable such that $t \in \left\{0,1\right\}$. We then convert \textit{v} into \textit{t} as follows:

\begin{equation}
\label{typecast-bv2bool}
t = \left\{ \begin{array}{ll}
v = k \rightarrow 0, \\
v \neq k \rightarrow 1 \end{array} \right.	
\end{equation}

\subsubsection{Arithmetic Overflow and Underflow}
\label{03-EncodingArithmeticOverflowandUnderflow}

Arithmetic overflow and underflow are frequent sources of bugs in embedded software. ANSI-C, like most programming languages, provides basic data types that have a bounded range defined by the number of bits allocated to each of them. Some model checkers (e.g., SMT-CBMC, F-Soft and Blast~\cite{Armando06,Ganai06,Blast08}) treat program variables either as unbounded integers or they do not generate VCs related to arithmetic overflow and consequently can produce false positive results when a VC cannot violate the boundary condition. In our work, we encode VCs related to arithmetic overflow and underflow in the following way: On arithmetic overflow of \emph{unsigned} integer types (e.g., \emph{unsigned int}, \emph{unsigned long int}), the ANSI-C standard requires that the result must be encoded as modulo (i.e., $r$ $mod$ $2$$^{w}$, where $r$ is the operation that caused overflow and $w$ is the width of the resulting type in terms of bits)~\cite{ISO99}. Hence, the result of this encoding is one greater than the largest value that can be represented by the resulting type. These semantics can be easily encoded using the background theories of the SMT solvers.

On the other hand, on arithmetic overflow of \emph{signed} types (e.g., \emph{int}, \emph{long int}), the ANSI-C standard does not define any behaviour to detect signed integer overflow and it only states that integer \textit{division-by-zero} must be detected. As a result, we consider arithmetic overflow on addition, subtraction, multiplication, division and negation operations. Formally, let $overflow$$^{*}$$(x, y)$ denote a literal that is true if and only if the multiplication of $x$ and $y$ is over LONG$\_$MAX and let $underflow$$^{*}$$(x, y)$ denote another literal that is true if and only if the multiplication of $x$ and $y$ is under LONG$\_$MIN. Let $res$$\_$$op$$^{*}$ be a literal that denotes the validity of the signed multiplication. Then, we add the following constraint:
\begin{equation}
\mathit{res\_op}^{*} \Leftrightarrow \neg \mathit{overflow}^{*}(x,y) \wedge \neg \mathit{underflow}^{*}(x,y) \nonumber
\end{equation}


The addition, subtraction and division are encoded in a similar way and are denoted by $\mathit{overflow}$$^{+}$, $\mathit{underflow}$$^{+}$, $\mathit{overflow}$$^{-}$, $\mathit{underflow}$$^{-}$, $\mathit{overflow}$$^{/}$. However, the function $\mathit{overflow}$$^{\sim}$$(x)$ takes only one argument and returns true if and only if the negation of $x$ is outside the interval given by LONG$\_$MIN and LONG$\_$MAX.

\subsubsection{Arrays}
\label{03-ArraysandStructuresUnions}

Arrays are encoded in a straight-forward manner using the domain theories, and we consider the \textit{WITH} operator and index operator $\left[\,\right]$ to be part of the encoding~\cite{Clarke04,Gries80}. These operators are mapped to the functions \textit{store} and \textit{select} of the array theory presented in Section~\ref{02-SatisfiabilityModuloTheories} respectively. For the \textit{with} operator, let \textit{a} be an array, \textit{i} be an integer variable, and \textit{v} be an expression with the type of the elements in \textit{a}. The operator \textit{with} takes \textit{a}, \textit{i}, and \textit{v} and returns an array that is exactly the same as array \textit{a} except that the value at index position \textit{i} is \textit{v} (if \textit{i} is within the array bounds). Formally, let $a'$ be $a \, \, with \left[i\right] := v,$ and \textit{j} an index of \textit{a}, then:
\begin{equation}
a'\left[j\right] = \left\{ \begin{array}{ll}
i = j \rightarrow v, \\
i \neq j \rightarrow a\left[j\right] \end{array} \right.	
\end{equation}

If an array index operation is out of bounds, the value of the index operator is a free variable, i.e., it is chosen non-deterministically. 

\subsubsection{Structures and Unions}
\label{03-StructuresUnions}

Structures and unions are encoded by using the theory of tuples in SMT and map update and access operations to the functions \textit{store} and \textit{select} of the tuples theory presented in Section~\ref{02-SatisfiabilityModuloTheories} respectively. As a result, we describe here only the encoding process of structures, but unions are encoded in a similar way. Let \textit{w} be a structure type, \textit{f} be a field name of this structure, and \textit{v} be an expression matching the type of the field \textit{f}. The expression \textit{store} takes \textit{w}, \textit{f}, and \textit{v} and returns a tuple that is exactly the same as tuple \textit{w} except that the value at field \textit{f} is \textit{v} and all other tuple elements remain the same. Formally, let $w'$ be $store(w,f,v)$ and \textit{j} be a field name of \textit{w}, then:
\begin{equation}
w'.j = \left\{ \begin{array}{ll}
j = f \rightarrow v, \\
j \neq f \rightarrow w.j \end{array} \right.	
\end{equation}

\subsubsection{Pointers}

The ANSI-C language offers two dereferencing operators $^{*}p$ and $p\left[i\right]$, where \textit{p} denotes a pointer (or array) and \textit{i} denotes an integer index. The front-end of CBMC removes all pointer dereferences bottom-up during the unwinding phase. Therefore, the ANSI-C pointers are treated as program variables and CBMC's VCG generates two properties related to \textit{pointer safety}: \textit{(i)} check if the pointer points to a correct object (represented by \emph{SAME$\_$OBJECT}) and \textit{(ii)} check if the pointer is neither \verb|NULL| nor an invalid object (represented by \emph{INVALID$\_$POINTER}).

We thus encode pointers using two fields of a tuple. Let \textit{p} denote the tuple which encodes a pointer type expression. The first field \textit{p.o}, encodes the object the pointer points to, while the second field \textit{p.i}, encodes an index within that object. It is important to note that in our encoding the field \textit{p.o} is dynamically adjusted in order to accommodate the object that the pointer points to. This approach is similar to the encoding of CBMC into propositional logic, but we use the background theories such as tuples and bit-vector arithmetic while CBMC encodes them by concatenating the bit-vectors.

Formally, let $p_{a}$ be a pointer expression that points to the object $a$ and $p_{b}$ be another pointer expression that points to the object $b$. Let $l_{s}$ be a literal and we then encode the property \emph{SAME$\_$OBJECT} by adding the following constraint:
\begin{equation}
l_{s} \Leftrightarrow \left(p_{a}.o = p_{b}.o\right)
\end{equation}

To check invalid pointers, the \verb|NULL| pointer is then encoded with an unique identifier denoted by $\eta$ and invalid object is denoted by $\nu$. Let $p$ denote a pointer expression. Hence we encode the property \emph{INVALID$\_$POINTER} by creating a literal $l_{i}$ and adding the following constraint:
\begin{equation}
l_{i} \Leftrightarrow \left(p.o \neq \nu\right) \wedge \left(p.i \neq \eta\right)
\end{equation}

%


It is important to note that in the case that a pointer points to single element of a scalar data type (e.g., \emph{int}, \emph{char}), then \textit{p.i} consists of 0 only. However, in case of an array consisting of elements of a scalar data type, \textit{p.i} is considered to be equal to the array index. As an example to explain our encoding, we modified the C program of Figure~\ref{fig:set-of-transformation}(a) so that a pointer \emph{p} points to the array \emph{a} as shown in line 3 of Figure~\ref{alg:C-program-with-pointer}. In addition to the constraints and properties shown in (\ref{quantifier-free-formulae-c}) and (\ref{quantifier-free-formulae-p}) (Section~\ref{sec:TheNewBackEndOfCBMC}), the front-end generates one additional constraint (i.e., the front-end treats the assignment \emph{p=a} in line 3 as \emph{p=$\&$a[0]}) and one additional VC (i.e., \emph{SAME$\_$OBJECT(p, $\&$a[0])}) for the C program of Figure~\ref{alg:C-program-with-pointer}. The constraint \emph{p$=$$\&$a[0]} is encoded as follows: the first element of the tuple (\textit{p.o}) contains the array \emph{a} and the second element (\textit{p.i}) contains the index whose value is equal to 0. In order to check the property specified by the \emph{assert} macro in line 8, we first add the value 2 to \textit{p.i} and then check whether \emph{p} and \emph{a} point to the same element. As \textit{p.i} exceeds the size of the object stored in \textit{p.o}, i.e., array \emph{a}, then the VC is violated and thus the \emph{assert} macro defined in line 8 is false.

\begin{figure}[ht]
\centering
\begin{lstlisting}
int main() {
  int a[2], i, x, *p;
  p=a;
 	if (x==0)
		a[i]=0;
	else
    a[i+2]=1;
	assert(*(p+2)==1);
}
\end{lstlisting}
\caption{C program with pointer to an \emph{array}.}
\label{alg:C-program-with-pointer}
\end{figure}

Structures consisting of \textit{n} fields with scalar data types are also manipulated like an array with \textit{n} elements. This means that the front-end of CBMC allows us to encode the structures by using the usual update and access operations. If the structure contains arrays, pointers and scalar data types, then \textit{p.i} points to the object within the structure only. As an example, Figure~\ref{alg:C-program-with-pointer-to-structure} shows a C program that contains a pointer to a \emph{struct} consisting of two fields (an array \emph{a} of integer and a \emph{char} variable \emph{b}). In order to reason about this C program, the front-end generates the constraints and properties and we then encode and pass the resulting formulae to the SMT solvers as $C$$\wedge$$\neg$$P$ (as shown in (\ref{quantifier-free-formulae-c-struct}) and (\ref{quantifier-free-formulae-p-struct})). 

As the \emph{struct} \emph{y} is declared as global in Figure~\ref{alg:C-program-with-pointer-to-structure} (lines 1-4), its members must be initialized before performing any operation as shown in (\ref{quantifier-free-formulae-c-struct}) (first line)~\cite{ISO99}. The assignment $p$ $=$ $\&y$ (line 7 of Figure~\ref{alg:C-program-with-pointer-to-structure}) is encoded by assigning the structure \emph{y} to the field \textit{p$_{1}$.o} and the value 0 to the field \textit{p$_{1}$.i}. However, the front-end does not generate any VC related to pointer safety since there is no violation of the pointer \emph{p} in the C program of Figure~\ref{alg:C-program-with-pointer-to-structure} (i.e., the pointer \emph{p} points to the correct object). As a result, the front-end performs static checking and does not generate unnecessary VCs. Thus, the pointer \emph{p} represented by the tuple \textit{p$_{1}$} is not used for reasoning about this program.

\begin{figure}[ht]
\centering
\begin{lstlisting}
struct x {
  int a[2];
  char b;
} y;
int main(void) {
  struct x *p;
  p=&y;
  p->a[1]=1;
  p->b='c';
  assert(p->a[1]==1);
  assert(p->b=='c');
}                   
\end{lstlisting}
\caption{C program with pointer to a \emph{struct}.}
\label{alg:C-program-with-pointer-to-structure}
\end{figure}

\begin{equation}
\label{quantifier-free-formulae-c-struct}
C := \left [ \begin{array}{ll} 
                y_{1} := store(store(y_{0}.a, 0, 0), 1, 0) \, \, \wedge \, \, y_{0}.b=0 \\ 
                \wedge \, \, p_{1}.o := y \, \, \wedge \, \, p_{1}.i := 0\\
                \wedge \, \, y_{2} := store(y_{1}, a, store(y_{1}.a, 0, 0)) \\
                \wedge \, \, y_{3} := store(y_{2}, a, store(y_{2}.a, 1, 1)) \\
                \wedge \, \, y_{4} := store(y_{3}, b, 99)
              \end{array} \right ]  \\ 
\end{equation}
\begin{equation}
\label{quantifier-free-formulae-p-struct}
P := \left [ \begin{array}{ll} 
                select(select(y_{4}, a), 1) =  1 \\ 
                \wedge \, \, select(y_{4}, b) =  99\\
              \end{array} \right ]  \\ 
\end{equation}

\section{Experimental Evaluation}
\label{05-experimental-results}

The experimental evaluation of our work consists of three parts. The first part in Section~\ref{sec:ComparisonOfSMTSolvers} contains the results of applying ESW-CBMC to the verification of fifteen ANSI-C programs using three different SMT solvers CVC3, Boolector and Z3. The purpose of this first part is thus to identify the most promising SMT solver for further development and experiments. CVC3, Boolector and Z3 are well suited for the purpose that they were written for and our intention is to integrate all of them into the back-end of CBMC, but firstly we need to prioritize the tasks. The second part, described in Section~\ref{sec:ComparisontoSMTCBMC}, contains the results of applying ESW-CBMC and SMT-CBMC to the verification of the official benchmark of the SMT-CBMC model checker. We use the official benchmark, because SMT-CBMC does not support some of the ANSI-C constructs commonly found in embedded software (e.g., bit operations, floating-point arithmetic, pointer arithmetic). As a result, the purpose of this second part is to evaluate ESW-CBMC's relative performance against SMT-CBMC. 

The third part in Section~\ref{sec:ComparisontoCBMC} contains the experimental results of applying CBMC and ESW-CBMC to the verification of embedded software used in telecommunications, control systems and medical devices. The purpose of this third part is to evaluate ESW-CBMC's relative performance against CBMC using standard embedded software benchmarks. All experiments were conducted on an otherwise idle Intel Xeon 5160, 3GHz server with 4 GB of RAM running Linux OS. For all benchmarks, the time limit has been set to 3600 seconds for each individual property. All times given are wall clock time in seconds as measured by the unix \textit{time} command through a single execution.

\subsection{Comparison of SMT solvers}
\label{sec:ComparisonOfSMTSolvers}

As a first step, we analyzed to which extent the SMT solvers support the domain theories that are required for SMT-based BMC of ANSI-C programs. For this purpose, we analyzed the following versions of the SMT solvers: CVC3 (1.5), Boolector (1.0) and Z3 (2.0). For the \textit{theory of linear and non-linear arithmetic}, Z3 and CVC3 do not support the remainder operator, but they allow us to define axioms to support it. Currently, Boolector does not support the theory of linear and non-linear arithmetic. In the \textit{theory of bit-vectors}, CVC3 does not support the division and remainder operators ($/$, \emph{rem}) for bit-vectors representing signed and unsigned integers. However, in all cases, axioms can be specified in order to improve the coverage. Z3 and Boolector support all word-level, bit-level, relational, arithmetic functions over unsigned and signed bit-vectors. In the theories of \textit{arrays} and \textit{tuples}, the verification problems only involve selecting and storing elements from/into arrays and tuples, respectively, and both domains thus comprise only two operations. These operations are fully supported by CVC3 and Z3, but Boolector does not support the theory of tuples.

\begin{table*}[t!]
\setlength{\tabcolsep}{3pt}
\begin{center} {\small
\begin{tabular}{|c|l|c|c|c||r|r|r|r|r|r|r|r|r|}
\hline
 & & & & & \multicolumn{3}{c|}{CVC3} & \multicolumn{3}{c|}{Boolector} & \multicolumn{3}{c|}{Z3} \\
 & Module & \multicolumn{1}{|c|}{$\#$L} & \multicolumn{1}{|c|}{B} & \multicolumn{1}{|c||}{$\#$P} & \multicolumn{1}{|c}{Size} & \multicolumn{1}{c}{Time} & \multicolumn{1}{c|}{Failed} & \multicolumn{1}{|c}{Size} & \multicolumn{1}{c}{Time} & \multicolumn{1}{c|}{Failed} & \multicolumn{1}{|c}{Size} & \multicolumn{1}{c}{Time} & \multicolumn{1}{c|}{Failed} \\
 \hline
1 & BubbleSort & 43 & 35  & 17 & 9031   & 28.27  & 0 & 3011    & 1.94   & 0 & 6057  & 2.03   & 0 \\
  &            & 43 & 140 & 17 & 146371 & MO & 1 & 48791   & 182.67 & 0 & 97722 & 163.15 & 0 \\
\hline
2 & SelectionSort & 34 & 35  & 17 & 6982 & 8.48 & 0 & 1955 & 0.78 & 0 & 5134 & 0.83 & 0 \\
  &               & 34 & 140 & 17 & 108832 & MO & 1 & 29885 & 74.59 & 0 & 79369 & 74.36 & 0 \\
\hline
3 & BellmanFord & 49 & 20 & 33 & 1076& 0.45 & 0 & 326 & 0.27 & 0 & 656 & 0.3 & 0 \\
\hline
4 & Prim & 79 & 8 & 30 & 4008 & 16.88 & 0 & 1296 & 0.5 & 0 & 3017 & 0.48 & 0\\
\hline
5 & StrCmp & 14 & 1000 & 6 & 9005 & 9.88 & 0 & 3003 & 91.145 & 0 & 7006 & 38.75 & 0\\
\hline
6 & SumArray & 12 & 1000 & 7 & 3001 & 1.22 & 0 & 1001 & 0.93 & 0 & 2003 & 4.74 & 0\\
\hline
7 & MinMax & 19 & 1000 & 9 & 17989 & MO & 1 & 5997 & 947.58 & 0 & 11994 & 6.22 & 0\\
\hline
8 & InsertionSort & 86 & 35  & 17 & 9337 & 35.57 & 0 & 3113 & 2.37 & 0 & 6328 & 2.51 & 0 \\
  &               & 86 & 140 & 17 & 147622 & MO & 1 & 49208 & TO & 1 & 98833 & 143 & 0 \\
\hline
9  & Fibonacci & 83 & 15 & 4 & 16 & 15.12 & 0 & 16 & 15.6 & 0 & 16 & 15.2 & 0 \\
\hline
10 & bs & 95 & 15 & 7 & 17 & 0.21 & 0 & 17 & 0.02 & 0 & 17 & 0.02 & 0 \\
\hline
11 & lms & 258 & 202 & 23 & 14810 & 1011.92 & 0 & 5005 & 138.74 & 0 & 10211 & 138.6 & 0 \\
\hline
12 & Cubic & 66 & 5 & 5 & 40 & 0.01 & 0 & 20 & 0.19 & 0 & 33 & 0.2 & 0 \\
\hline
13 & BitWise & 18 & 8 & 1 & 77 & 272.38 & 0 & 27 & 7.51 & 0 & 53 & 28.37 & 0 \\
\hline
14 & adpcm$\_$encode & 149 & 41 & 12 & 6417 & 211.81 & 0 & 2377 & 738.86 & 0 & 4878 & 5.49 & 0 \\
15 & adpcm$\_$decode & 111 & 41 & 10 & 23885 & 43.77 & 0 & 9121 & 20.16 & 0 & 19270 & 14.31 & 0 \\
\hline
\end{tabular} }
\end{center}
\caption{Results of the comparison between CVC3, Boolector and Z3. Time-outs are represented with TO in the Time column; Examples that exceed available memory are represented with MO in the Time column.}
\label{table:results-of-the-benchmark-part-i}
\label{turns}
\end{table*}

In order to evaluate the SMT solvers, we used a number of ANSI-C programs taken from standard benchmark suites. The results of this first part are shown in Table~\ref{table:results-of-the-benchmark-part-i}. The first seven programs are taken from the benchmark suite of the SMT-CBMC model checker~\cite{Armando06}. These programs depend on a positive integer \textit{N} that defines the size of the arrays in the programs and/or the number of iterations done by the program. Armando et al. already proved that this class of programs allows us to assess the scalability of the model checking tools on problems of increasing complexity~\cite{Armando06}. The next four programs are taken from the SNU Real-Time benchmarks suite~\cite{SNURealTime09}. These programs implement the insertion sort algorithm, Fibonacci function, binary search algorithm and the least mean-square (LMS) adaptive signal enhancement. Program 9 is taken from the MiBench benchmark and implements the root computation of cubic equations. Program 10 is taken from the CBMC manual~\cite{Clarke04} and implements the multiplication of two numbers using bit operations. The last two programs are taken from the High Level Synthesis benchmarks suite~\cite{HighLevelSynthesis09} and implement the encoder and decoder of the adaptive differential pulse code modulation (ADPCM). The C programs from 8 to 15 contain typical ANSI-C constructs found in embedded software, i.e., they contain linear and non-linear arithmetic and make heavy use of bit operations.

Table~\ref{table:results-of-the-benchmark-part-i} shows the results of the comparison between CVC3, Boolector and Z3. The first column $\#$L gives the total number of lines of code, the second column B gives the unwinding bound while the third column $\#$P gives the number of properties to be verified for each ANSI-C program. \textit{Size} gives the total number of variables that are needed to encode the constraints and properties of the ANSI-C programs. \textit{Time} provides the average time in seconds to check all properties of a given ANSI-C program and \textit{Failed} indicates how many properties failed during the verification process. Here, properties can fail for two reasons: either due to a time out (TO) or due to memory out (MO). As we can see in Table~\ref{table:results-of-the-benchmark-part-i}, Z3 runs slightly faster than Boolector and CVC3 except for the ANSI-C programs \textit{StrCmp} and \textit{SumArray}. As we mentioned previously, the purpose of this evaluation is to prioritize the integration of the SMT solvers into the back-end of CBMC and not to define the best SMT solver. Since Z3 supports most of the occurring operations, we chose to continue the development with Z3.

\subsection{Comparison to SMT-CBMC}
\label{sec:ComparisontoSMTCBMC}

This subsection describes the evaluation of ESW-CBMC against another SMT-based BMC that was developed in~\cite{Armando06,Armando09}. In order to carry out this evaluation, we took the official benchmark of SMT-CBMC tool available at~\cite{smtcbmc09}. SMT-CBMC has been invoked by setting manually the file name and the unwinding bound (i.e., SMT-CBMC {\tt -file name} {\tt -bound n}). Furthermore, we used the default solver of SMT-CBMC (i.e, CVC3 1.5) against the default solver of ESW-CBMC (i.e., Z3 2.0) as well as ESW-CBMC connected to CVC3 1.5. Table~\ref{table:results-of-the-benchmark-part-ii} shows the results of this evaluation. 

\begin{table}[t!]
\centering {\small
\begin{tabular}{|l|c|c||r|r|r|}
\hline
 & \multicolumn{1}{|l|}{} & \multicolumn{1}{c||}{} & \multicolumn{2}{c|}{ESW-CBMC} & \multicolumn{1}{c|}{SMT-CBMC} \\
 \multicolumn{1}{|l|}{Module} & \multicolumn{1}{|c|}{$\#$L} & \multicolumn{1}{c||}{B} & \multicolumn{1}{c}{Z3} & \multicolumn{1}{c|}{CVC3} & \multicolumn{1}{c|}{CVC3}\\\hline
BubbleSort & 43 & 35 & 2.03 & 28.27 & 94.5\\
           & 43 & 140 & 163.15 & MO & $\ast$  \\
\hline
SelectionSort & 34 & 35  & 0.83 & 8.48 & 66.52\\
              & 34 & 140 & 74.36 & MO & MO\\
\hline
BellmanFord & 49 & 20 & 0.3 & 0.45 & 13.62\\
\hline
Prim & 79 & 8 & 0.48 & 16.88 & 18.36\\
\hline
StrCmp & 14 & 1000 & 38.75 & 9.88 & TO\\
\hline
SumArray & 12 & 1000 & 4.74 & 1.22 & 113.8\\
\hline
MinMax & 19 & 1000 & 6.22 & MO & MO\\\hline \end{tabular} }
\caption{Results of the comparison between ESW-CBMC and SMT-CBMC.}
\label{table:results-of-the-benchmark-part-ii}
\end{table}

If CVC3 is used as SMT solver, both tools run out of memory (although only after exceeding the time out) and fail (due to many dynamic choice points represented by $^{\ast}$) to analyze \textit{BubbleSort} and \textit{SelectionSort} for large \textit{N} (\textit{N}=140), and \textit{MinMax}. This indicates some problems in the solver itself, rather than in verification tools. In addition, SMT-CBMC runs out of time to analyze the program \textit{StrCmp}. However, if Z3 is used as solver for ESW-CBMC, the difference becomes even more noticeable and ESW-CBMC outperforms SMT-CBMC consistently by a factor of 20-40.


\subsection{Comparison to CBMC}
\label{sec:ComparisontoCBMC}

In order to evaluate ESW-CBMC's relative performance against CBMC, we analyze different benchmarks such as SNU Real-Time, PowerStone, NEC and NXP~\cite{SNURealTime09,PowerStone98,necla09,nxp09}. The SNU Real-Time benchmarks contain ANSI-C programs that implement cyclic redundancy check, Fast Fourier Transform, LMS adaptive signal enhancement, JPEG, matrix multiplication, LU decomposition and root computation of quadratic equations. The PowerStone benchmarks contain graphics applications, ADPCM encoder and decoder, paging communication protocols and bit shifting applications. The NEC benchmark contains an implementation of the Laplace transform. The NXP benchmarks are taken from the set-top box of NXP semiconductors that is used in high definition internet protocol (IP) and hybrid digital TV (DTV) applications. The embedded software of this platform relies on the Linux operating system and makes use of different applications such as \textit{(i)} \textit{LinuxDVB} that is responsible for controlling the front-end, tuners and multiplexers, \textit{(ii)} \textit{DirectFB} that provides graphics applications and input device handling and \textit{(iii)} \textit{ALSA} that is used to control the audio applications. This platform contains two embedded processors that exchange messages via an inter-process communication (IPC) mechanism.
 
We evaluated CBMC version 2.9 and we invoke both tools (i.e., CBMC and ESW-CBMC) by setting manually the file name, the unwinding bound and the overflow check (i.e., CBMC {\tt file} {\tt --unwind n} {\tt --overflow-check}). Table~\ref{table:results-of-the-benchmark-part-iii} shows the results when applying CBMC and ESW-CBMC to the verification of the embedded software benchmarks.
  
\begin{table*}[t]
\setlength{\tabcolsep}{3pt}
\begin{center} {\small
\begin{tabular}{|c|l|c|c|c||r|r|r|r|r|r||r|r|r|r|r|r|}
\hline
 &  & & & & \multicolumn{6}{c||}{CBMC} & \multicolumn{6}{c|}{ESW-CBMC} \\  \cline{6-17}
 & & & & & \multicolumn{3}{c|}{Time} & \multicolumn{3}{c||}{$\#$P} & \multicolumn{3}{c|}{Time} & \multicolumn{3}{c|}{$\#$P} \\ \cline{6-17}
 & Module & $\#$L & B & $\#$P & \multicolumn{1}{c|}{\rotatebox{90}{Encoding}} & \multicolumn{1}{c|}{\rotatebox{90}{Decision} \rotatebox{90}{Procedure}} & \multicolumn{1}{c|}{\rotatebox{90}{Total}} & \multicolumn{1}{c|}{\rotatebox{90}{Passed}} & \multicolumn{1}{c|}{\rotatebox{90}{Violated}} & \multicolumn{1}{c||}{\rotatebox{90}{Failed}} & \multicolumn{1}{c|}{\rotatebox{90}{Encoding}} & \multicolumn{1}{c|}{\rotatebox{90}{Decision} \rotatebox{90}{Procedure}} & \multicolumn{1}{c|}{\rotatebox{90}{Total}} & \multicolumn{1}{c|}{\rotatebox{90}{Passed}} & \multicolumn{1}{c|}{\rotatebox{90}{Violated}} & \multicolumn{1}{c|}{\rotatebox{90}{Failed}} \\\hline
1 & sensor & 603 & 5 & 167 & 2.04 & 0.002 & 2.04 & 167 & 0 & 0 & 1.23 & 0.02 & 1.26 & 167 & 0 & 0\\
\hline
2 & crc & 125 & 257 & 18 & 5.60 & 0.003 & 5.60 & 18 & 0 & 0 & 4.08 & 0.07 & 4.16 & 18 & 0 & 0\\
\hline
3 & fft1 & 218 & 9 & 72 & 0.44 & 0.001 & 0.44 & 72 & 0 & 0 & 0.43 & 0.005 & 0.43 & 72 & 0 & 0\\
\hline
4 & fft1k & 155 & 1025 & 39 & MO & MO & MO & 0 & 0 & 39 & 2337.83 & 0.055 & 2337.88 & 39 & 0 & 0\\
\hline 
5 & fibcall & 83 & 30 & 2 & 0.19 & 0 & 0.19 & 2 & 0 & 0 & 0.15 & 0.002 & 0.15 & 2 & 0 & 0\\
\hline 
6 & fir & 314 & 34 & 25 & 4.88 & 0.02 & 4.9 & 25 & 0 & 0 & 3.36 & 0.68 & 4.04 & 25 & 0 & 0\\
\hline 
7 & insertsort & 86 & 10 & 17 & 0.36 & 0.005 & 0.37 & 17 & 0 & 0 & 0.31 & 0.02 & 0.32 & 17 & 0 & 0\\
\hline 
8 & jfdctint & 374 & 65 & 331 & 1.22 & 0.001 & 1.22 & 330 & 1 & 0 & 0.45 & 2.41 & 2.86 & 330 & 1 & 0\\
\hline 
9 & lms & 258 & 202 & 35 & MO & MO & MO & 0 & 0 & 35 & 132.6 & 0.24 & 132.84 & 35 & 0 & 0\\
\hline 
10 & ludcmp & 144 & 7 & 88 & 4.52 & TO & TO & 87 & 0 & 1 & 0.017 & 1.44 & 1.46 & 88 & 0 & 0\\
\hline 
11 & matmul & 81 & 6 & 31 & 1.16 & 0 & 1.16 & 31 & 0 & 0 & 1.06 & 0.012 & 1.07 & 31 & 0 & 0\\
\hline 
12 & qurt & 164 & 20 & 8 & 18.83 & TO & TO & 7 & 0 & 1 & 1.22 & 7.7 & 8.92 & 8 & 0 & 0\\
\hline 
13 & bcnt & 86 & 17 & 162 & 4.42 & 0.05 & 4.47 & 162 & 0 & 0 & 1.24 & 0.89 & 2.13 & 162 & 0 & 0\\
\hline
14 & blit & 95 & 1 & 129 & 0.21 & 0.001 & 0.21 & 128 & 1 & 0 & 0.13 & 0.28 & 0.41 & 128 & 1 & 0\\
\hline 
15 & pocsag & 521 & 42 & 183 & 15.32 & 0.1 & 15.42 & 182 & 1 & 0 & 12.33 & 5.77 & 18.1 & 182 & 1 & 0\\
\hline 
16 & adpcm & 473 & 100 & 553 & 74.34 & 3.52 & 77.86 & 553 & 0 & 0 & 45.73 & 9.24 & 54.97 & 553 & 0 & 0\\
\hline 
17 & laplace & 110 & 11 & 76 & 30.81 & TO & TO & 0 & 0 & 76 & 12.32 & 0.29 & 12.62 & 76 & 0 & 0\\
\hline 
18 & exStbKey & 558 & 20 & 18 & 1.23 & 0.002 & 1.23 & 18 & 0 & 0 & 1.22 & 0.004 & 1.23 & 18 & 0 & 0\\
\hline 
19 & exStbHDMI & 1045 & 15 & 25 & 167.91 & 78.97 & 246.88 & 25 & 0 & 0 & 164.43 & 33.53 & 197.96 & 25 & 0 & 0\\
\hline 
20 & exStbLED & 430 & 40 & 6 & 195.97 & 129.8 & 325.77 & 6 & 0 & 0 & 165.63 & 44.53 & 210.16 & 6 & 0 & 0\\
\hline 
21 & exStbHwAcc & 1432 & 1000 & 113 & 0.67 & 0.002 & 0.67 & 113 & 0 & 0 & 0.72 & 0.004 & 0.73 & 113 & 0 & 0\\
\hline 
22 & exStbResolution & 353 & 200 & 40 & 271.8 & 319.13 & 590.93 & 40 & 0 & 0 & 269.31 & 1161.16 & 1430.47 & 40 & 0 & 0\\
\hline 
\end{tabular} }
\end{center}
\caption{Results of the comparison between CBMC and ESW-CBMC}
\label{table:results-of-the-benchmark-part-iii}
\label{turns}
\end{table*}

As we can see in Table~\ref{table:results-of-the-benchmark-part-iii}, CBMC is not able to check the programs \textit{fft1k} and \textit{lms} due to memory limitations. Moreover, CBMC takes considerably more time than ESW-CBMC to model check the programs \textit{ludcmp}, \textit{qurt} and \textit{laplace}. In addition, ESW-CBMC runs faster than CBMC for the programs \textit{adpcm}, \textit{exStbHDMI} and \textit{exStbLED}. The only case that CBMC runs faster than ESW-CBMC is with the program \textit{exStbResolution}. For the remaining benchmarks, the verification times of ESW-CBMC and CBMC are very close. It is important to point out that the encoding time of ESW-CBMC, for all analyzed programs, is slightly faster than the encoding time of CBMC. The results in Table~\ref{table:results-of-the-benchmark-part-iii} allow us to assess quantitatively that ESW-CBMC scales significantly better than CBMC for problems that involve tight interplay between non-linear arithmetic, bit operations, pointers and array manipulations. In addition, both tools were able to find undiscovered bugs related to arithmetic overflow, invalid pointer and pointer arithmetic in the programs \textit{jfdctint}, \textit{blit} and \textit{pocsag} respectively.

\section{Related Work}
\label{06-related-work}

SMT-based BMC is gaining popularity in the formal verification community due to the advent of sophisticated SMT solvers built over efficient SAT solvers~\cite{CVC07,Boolector09,Z08}. Previous work related to SMT-based BMC~\cite{Armando06,Ganai06,Xu08,Armando09} combined decision procedures for the theories of uninterpreted functions, arrays and linear arithmetic only, but did not encode key constructs of the ANSI-C programming language such as bit operations, floating-point arithmetic and pointers. Ganai and Gupta describe a verification framework for BMC which extracts high-level design information from an extended finite state machine (EFSM) and applies several techniques to simplify the BMC problem~\cite{Ganai06,Ganai08}. However, the authors flatten the structures and arrays into scalar variables in such a way that they use only the theory of integer and real arithmetic in order to solve the verification problems that come out in BMC.

Armando et al. also propose a BMC approach using SMT solvers for C programs~\cite{Armando06,Armando09}. However, they only make use of linear arithmetic (addition and multiplication by constants), arrays, records and bit-vectors in order to solve the verification problems. As a consequence, their SMT-CBMC prototype does not address important constructs of the ANSI-C programming language such as non-linear arithmetic and bit-shift operations. Xu proposes the use of SMT-based BMC to verify real-time systems by using TCTL to specify the properties~\cite{Xu08}. The author considers an informal specification (written in English) of the real-time system and then models the variables using integers and reals and represents the clock constraints using linear arithmetic expressions.

De Moura et al. present a bounded model checker that combines propositional SAT solvers with domain-specific theorem provers over infinite domains~\cite{Moura02}. Differently from other related work, the authors abstract the Boolean formula and then apply a lazy approach to refine it in an incremental way. This approach is applied to verify timed automata and RTL level descriptions. Jackson et al.~\cite{Jackson08} discharge several verification conditions from programs written in the Spark language to the SMT solvers CVC3 and Yices as well as to the theorem prover Simplify. The idea of this work is to replace the Praxis prover by CVC3, Yices and Simplify in order to generate counter-example witnesses to verification conditions that are not valid. This is an ongoing project and several improvements are planned to be integrated into their tool.

Recently, a number of static checkers have been developed in order to trade off scalability and precision. Calysto is an efficient static checker that is able to verify VCs related to arithmetic overflow, null-pointer dereferences and assertions specified by the user~\cite{bh08calysto}. The VCs are passed to the SMT solver SPEAR which supports boolean logic, bit-vector arithmetic and is highly customized for the VCs generated by Calysto. However, Calysto does not support float-point operations and unsoundly approximates loops by unrolling them only once. As a consequence, soundness is relinquished for performance. Saturn is another efficient static checker that scales to larger systems, but with the drawback of losing precision by supporting only the most common integer operators and performing at most two unwindings of each loop~\cite{saturn05}.

\section{Conclusions}
\label{07-conclusion}


In this work, we have investigated SMT-based verification of ANSI-C programs, in particular embedded software. We have described a new set of encodings that allow us to reason accurately about bit operations, unions, float-point arithmetic, pointers and pointer arithmetic and we have also improved the performance of SMT-based BMC for embedded software by making use of high-level information to simplify the unrolled formula. Our experiments constitute, to the best of our knowledge, the first substantial evaluation of this approach over industrial applications. The results show that our approach outperforms CBMC~\cite{Clarke04} and SMT-CBMC~\cite{Armando06} if we consider the verification of embedded software. SMT-CBMC still has limitations not only in the verification time (due to the lack of simplification based on high-level information), but also in the encodings of important ANSI-C constructs used in embedded software. CBMC is a bounded model checker for full ANSI-C, but it has limitations due to the fact that the size of the propositional formulae increases significantly in the presence of large data-paths and high-level information is lost when the verification conditions are converted into propositional logic (preventing potential optimizations to reduce the state space to be explored). For future work, we intend to investigate the application of termination analysis~\cite{Cook06} and incorporate reduction methods to simplify the \textit{k}-model.

\small{\paragraph{\textbf{Acknowledgement}} We thank D. Kroening, C. Wintersteiger and L. Platania for many helpful discussions about CBMC and SMT-CBMC model checking tools. We also thank L. de Moura and R. Brummayer for analyzing the VCs generated by ESW-CBMC and for indicating the most suitable configuration parameters and encoding for the SMT solvers Z3 and Boolector respectively. We also thank D. Kroening and J. Colley for reviewing a draft version of this paper.}



%




\end{document}